\documentclass[aps,prb,twocolumn,showpacs,preprintnumbers,amsmath,amssymb]{revtex4}
\usepackage{graphicx}
\usepackage{hyperref,textcomp}
\usepackage{natbib}
\usepackage{dcolumn}
\usepackage{bm}

\begin{document}

\title{Chemical functionalization of graphene}

\author{D. W. Boukhvalov}
\email{D.Bukhvalov@science.ru.nl} \affiliation{Institute for
Molecules and Materials, Radboud University of Nijmegen, NL-6525
ED Nijmegen, the Netherlands}
\author{M. I. Katsnelson}
\affiliation{Institute for Molecules and Materials, Radboud
University of Nijmegen, NL-6525 ED Nijmegen, the Netherlands}

\date{\today}

\begin{abstract}
Experimental and theoretical results on chemical functionalization
of graphene are reviewed. Using hydrogenated graphene as a model
system, general principles of the chemical functionalization are
formulated and discussed. It is shown that, as a rule, 100\%
coverage of graphene by complex functional groups (in contrast
with hydrogen and fluorine) is unreachable. A possible destruction
of graphene nanoribbons by fluorine is considered. The
functionalization of infinite graphene and graphene nanoribbons by
oxygen and by hydrofluoric acid is simulated step by step.
\end{abstract}



\maketitle

\section{Introduction}

Chemical functionalziation is one of the main methods to
manipulate the physical and chemical properties of nanoobjects
\cite{au1,C60chem1,C60chem2,NTchem,noncov,bioNT,
cntf1,cntf2,cntf3,cntf4} and to study mechanisms of interaction of
the nanoobjects with their environment (e.g., to investigate the
stability of desired properties with respect to oxidation,
hydrogenation, etc.). Initially, graphene was used by
theoreticians as a simple model to describe properties of
nanotubes \cite{Duplock,Leht,Dragan,expand,XPS,ahuja1,h2so4}.
After the discovery of graphene \cite{first} and of its
extraordinary electronic properties \cite{r1,r2,r3,r4} the
chemical functionalization of graphene became a focus of especial
interest in contemporary chemical physics. The main motivations of
these studies are (i) modification of electronic properties via
opening the energy gap in the electron spectrum of single- and
bilayer graphene \cite{Sofo,H,bilay,Peres,CrO3,BN, Hsubmit}; (ii)
potential use of graphene for hydrogen storage
\cite{Sofo,H,Li,Ti,Ataca,Yakobson,me2}; (iii) decoration of
various defects in graphene \cite{Duplock,swss,jacsvac,N,B,SW};
(iv) a way to make graphene magnetic for potential use in
spintronics \cite{jap1,jap2,H,Yazyev,SW,nmagn}; (v) search for
ways to produce ``cheap graphene'' by chemical reduction of
graphite oxide and manipulation of its electronic
\cite{GO,Si,ultra} and mechanical \cite{polymer,Schatz,Car1}
properties; (vi) the functionalization of graphene edges in
graphene nanoribbons
\cite{Louie,Kaxiras,YK,Yang,Ferarri2,Hod1,Hod2,Hod3,Hod4,Huang,magnonano,
kit1,kit2,O2,tranz1,tranz2,White,Mauri,fin} and their protection
\cite{SW}; (vii) oxidation and cracking of graphene \cite{unzip}
as tools to create graphene nanostructures of a given shape
\cite{cut1,cut2,cut3}.

There are two main types of functionalization, namely, a covalent
one, with the covalent bond formation, and non-covalent, due to
the van der Waals forces only. Most of the works deal with the
covalent functionalization, and only few of them
\cite{Peres,CrO3,Tim2,Tim3,antwerp, antwerp2,moldop} treat the
non-covalent one which is, of course, not surprising. First, the
covalent functionalziation results in much stronger modification
of geometric and electronic structures of graphene. Second, the
most of the density functional codes used now do not allow to
include the effects of the van der Waals interactions
\cite{vdW1,vdW2,vdW3} playing a crucial role in the non-covalent
functionalization. In those cases the local density approximation
(LDA) \cite{pz} is mostly used, instead of the standard
generalized gradient approximation (GGA) \cite{PBE} which gives
usually rather good results for layer compounds, such as graphite,
hexagonal boron nitride and MoS$_2$, due to error cancellation
\cite{FLAPW}.

Sometimes, the chemical functionalziation in graphene is related
to the ionic chemical bond. There are the cases of graphene layers
in graphite intercalated compounds \cite{GIC1,GIC2} and graphene
layers at metal surfaces
\cite{Ni1,Ni10,Ni2,Ni3,Ni4,Ru1,Ru2,Ir,Pt,me1,me3,me4}. This type
of functionalization is important to study possible
superconductivity in graphene, by analogy with the superconducting
intercalates CaC$_6$ and YbC$_6$ \cite{CaC6exp,CaC6th}.

Here we restrict ourselves only by the case of covalent
functionalization of graphene. We will start with the
hydrogenation of graphene, as a prototype of chemical
functionazliation (Section 3). In Section 4, other examples will
be considered, including realistic models of the functionalization
by diatomic molecules, such as O$_2$ and HF. In Section 5 we will
discuss effects of finite width on the functionalization of the
graphene nanoribbons (GNR). We will give a resume of the work and
discuss some perspectives in Section 5.

\section{Computational method}

 Our calculations have been performed with the SIESTA
code \cite{siesta1, siesta2, siesta3, siesta4} using the
generalized gradient approximation (GGA) \cite{PBE} to DFT and
Troullier-Martins \cite{TM} pseudopotentials. We used energy mesh
cutoff 400 Ry, and $k$-point mesh in Monkhorst-Park scheme
\cite{MP}. During the optimization, the electronic ground states
was found self-consistently by using norm-conserving
pseudopotentials to cores and a double-$\zeta$ plus polarization
basis of localized orbitals for carbon, oxygen and fluorine, and
double-$\zeta$ one for hydrogen. Optimization of the bond lengths
and total energies was performed with an accuracy 0.04 eV /\AA
~and 1 meV, respectively.

In our study of geometric distortions induced by the chemisorption
of single hydrogen atom which will be described below (Section 3)
a supercell containing 128 carbon atoms with $k$-point
4$\times$4$\times$1 mesh is used. For the cases of maximal
coverage of graphene by various chemical species (Section 4) the
standard elementary cell of graphene with two carbon atoms is
used, similar to the case of full hydrogenation \cite{H}, with
$k$-point 20$\times$20$\times$2 mesh. To simulate interaction of
bulk graphene with hydrofluoric acid the supercell with 8 carbon
atoms is used, similar to our previous works \cite{H,GO,bilay}
with $k$-point 11$\times$11$\times$1 mesh. To investigate chemical
functionalization of graphene nanoribbons (Section 5) zigzag
stripes with the width 22 and 66 carbon atoms are used, similar to
Ref. \cite{SW}, with 1$\times$13$\times$1 $k$-mesh.

For each step of the functionalization, its energy is defined as
E$_{chem_n}$ = E$_n$ - E$_{n-1}$ - E$_{ads}$, where E$_n$ and
E$_{n-1}$  are the total energies of the system for n-th and
n-1-th steps and E$_{ads}$ is the total energy of the adsorbed
molecule. In contrast with the standard definition of the
chemisorption which we used in the previous works, this definition
allows to evaluate energy favor or disfavor of each step
separately. We choose E$_{ads}$ as a total energy of the
corresponding molecule in gaseous phase. This is the most natural
choice to estimate stability of chemically modified graphene. This
is the crucial issue, for example, for the case of graphite oxide
where the desorption of hydroxy groups in the main obstacle to
derive a ``cheap graphene'' \cite{GO}. Note that molecular oxygen
exists in both triplet (spin-polarized) and singlet
(non-spin-polarized) states, the former being more energetically
favorable by 0.98 eV (we have found for this energy difference the
value 1.12 eV). We will use here, as well as in the previous work
\cite{GO} the energy of singlet O$_2$; to obtain the data with
respect to magnetic oxygen one needs merely to shift up the
presented energies by 1.12 eV. When considering oxidation of
nanotubes both energy of singlet \cite{O1,O2a,O4,O5} 
and triplet \cite{O3,O4,O5} oxygen are used.

\section{Hydrogen on graphene}

\begin{figure}
\includegraphics[width=3.2 in]{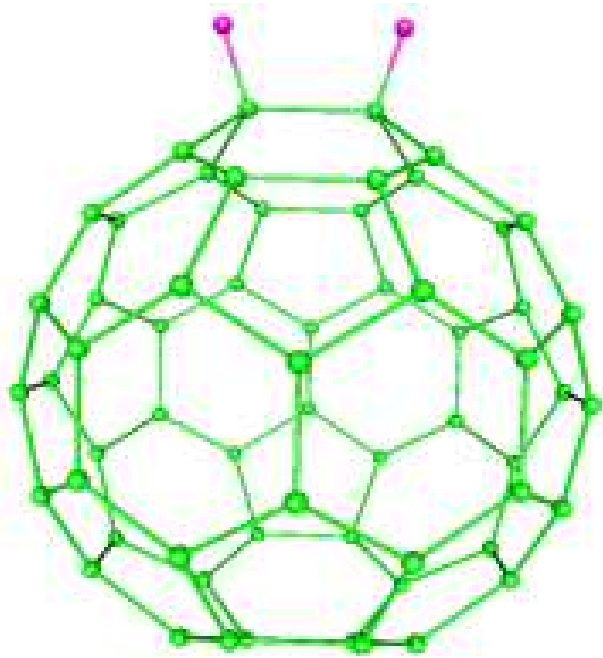}
\caption{(color online) Optimized geometric structure of
C$_{60}$H$_2$} \label{C60H2}
\end{figure}

Hydrogenation of carbon nanoobjects is a subject of special
interest starting from the discovery of fullerenes \cite{C60}, due
to potential relevance of this issue for the hydrogen storage
problem and a general scientific importance for chemistry. Soon
after the synthesis of the first fullerenes theoretical
\cite{C60H2th1,C60H2th2} and experimental \cite{C60H2exp} works
appeared on the minimal hydrogenated fullerene, C$_{60}$H$_2$. It
was shown in these works that the most stable configuration of
hydrogen on the fullerene corresponds to the functionalization of
the neighboring carbon atoms (1,2, according to chemical
terminology, see Fig. \ref{C60H2}). Further investigations result
in discovery of numerous systems with larger contents of hydrogen,
up to C$_{60}$H$_{36}$ \cite{C60H36}. Further hydrogenation of the
fullerenes turns out to be impossible, due to their specific
geometric structure. It is important that all known stable
compounds C$_{60}$H$_x$ contain even numbers of the hydrogen
atoms. Together with other known facts \cite{C60chem1,C60chem2} it
allowed to formulate the main principle for the hydrogenation of
the fullerenes: hydrogen atoms can be bonded either with the
neighboring or with the opposite atoms in carbon hexagons (1,2 or
1,4, in the chemical terminology). Another issue of special
interest was magnetism of hydrogenated fullerenes caused by
unpaired electrons on broken bonds. This can be realized only for
odd numbers of hydrogen atoms per buckyball \cite{C60Hmagn}.
Unfortunately, these hydrogenated fullerenes are unstable and
observed only as intermediate products of some chemical reactions;
for even numbers of the hydrogen atoms magnetism does not survive
and disappears with a time \cite{C60H24}. It is worth to note that
the hydrogenated fullerenes themselves are quite stable chemically
and do not loose hydrogen even under the action of high pressures
\cite{C60H36press}.

The discovery of single-wall carbon nanotubes \cite{NT,NT2}
(further we will discuss only this kind of the nanotubes) and
successful attempts of their reversible hydrogenation
\cite{Dillon1,Dillon2}, which makes the nanotubes more prospective
materials for hydrogen storage than fullerenes, have inspired
numerous theoretical models of the nanotube hydrogenation.
Geometric structure of the nanotubes differs essentially from that
of the fullerenes. Single-wall carbon nanotubes can be represented
as a graphene scroll of length up to few microns and width 1-1.5
nm.

Importantly, armchair and zigzag nanotubes exist with different
structure of the honeycomb net on their surface, and the nanotubes
can be chiral which is essential for their chemical
functionalization \cite{NTchem}. To avoid these complications,
graphene is frequently used to simulate chemical functionalization
of the nanotubes \cite{Duplock,Leht,Dragan,expand,XPS}. However,
as was discussed in Ref. \cite{H}, a curvature radius of typical
nanotubes (1.0 to 1.5 nm) is comparable with the radius of
geometric distortions induced by hydrogen. Effects of the
curvature of nanotubes on the hydrogenation was studied in detail
recently \cite{KrashNJP} and appeared to be quite essential, the
hydrogenation being different in two directions, along and around
the nanotubes. Influence of different types of chirality on
chemistry of nanotubes and differences with graphene was discussed
also in Ref. \cite{Tomanek,ahuja2}.

Let us discuss briefly the main results of these works. Potential
barriers for the cases of adsorption of a single hydrogen atom on
graphene and several types of nanotubes, as well as formation of
hydrogen chains on graphene and around the nanotubes have been
studied in Ref. \cite{Dragan}. Modification of the electronic
structure of perfect graphene and graphene with the Stone-Wales
defect at the adsorption of single hydrogen atom was discussed in
Ref. \cite{Duplock}. The authors \cite{Leht} investigated a
relation of magnetism with chemisorption of hydrogen for perfect
graphene and for two types of defects, that is, monovacancy and
interlayer carbon atom in graphite. Sluiter and Kawazoe
\cite{expand} used the cluster expansion algorithm to find the
most stable configuration for complete coverage of graphene by
hydrogen, to model the hydrogenation of nanotubes. This
configuration corresponds to so called graphane which has been
studied afterwards by the DFT calculations \cite{Sofo,H}. Wessely
et al \cite{XPS} simulated the core level spectra of carbon for
the case of chemisorption of single hydrogen atom on graphene, to
interpret existing experimental data on hydrogenated nanotubes.

Graphene was used also to model the hydrogenation of graphite
\cite{HOPG,metastab,Ferro1,Yazyev3}. After the discovery of
graphene several works have been carried out on the hydrogenation
of graphene itself, to modify its physical properties. In Ref.
\cite{Yazyev} the chemisorption of single hydrogen atom was
studied in more detail than before, as well as interactions of
magnetic moments arising, due to unpaired electrons, at the
hydrogenation of carbon atoms belonging to the same sublattice (it
was shown that in this case the interaction turns out to be
ferromagnetic). At the same time, according to Ref. \cite{Yazyev}
the combination of three hydrogen atoms (two of them belonging to
sublattice A and the third one to sublattice B) remains magnetic
which may be important in light of discussions of potential
magnetism of carbon systems.

\begin{figure}
\rotatebox{-90}{
\includegraphics[height=3.2 in]{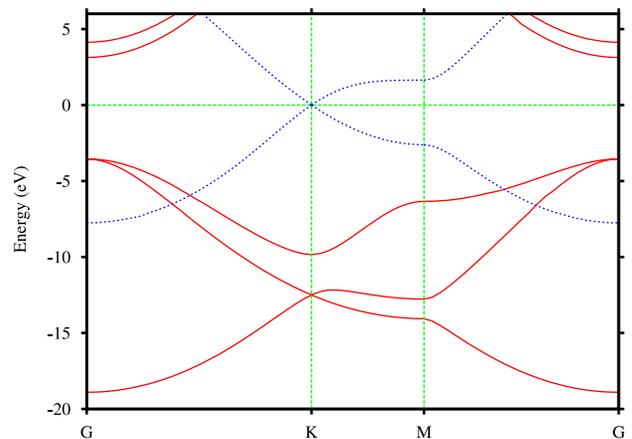}}
\caption{\label{bands}(color online) Band structure of a single
graphene layer. Solid red lines are $\sigma$ bands and dotted blue
lines are $\pi$ bands.}
\end{figure}

\begin{figure}
\includegraphics[width=3.2 in]{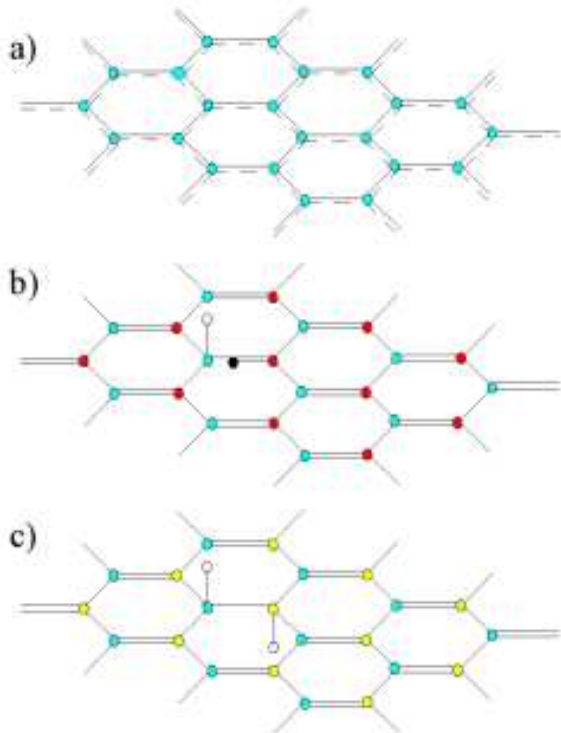}
\caption{A schematic representation of chemical bonds in (a) pure
graphene (dashed lines correspond to delocalized double bands),
(b) graphene with single chemisorbed hydrogen atom (white circle,
the black circle corresponds to unpaired electron at the broken
bond, it distributes in the sublattice shown in red), (c) graphene
with chemisorbed pair of hydrogen atoms (nonequivalent sublattices
are shown in different colors). \label{sublatt} (color online)}
\end{figure}

Sofo and coworkers \cite{Sofo} considered theoretically a
hypothetic material, graphane, that is, graphite with completely
hydrogenated carbon layers (in each layer, all hydrogen atoms
coupled with one sublattice are situated above, and with another
sublattice - below the layer). This structure corresponds to
weakly coupled diamond-like layers, with sp$^3$ hybridization,
instead of sp$^2$ in graphite. This change of hybridization
results in opening a gap of order 3 eV in electron energy
spectrum, due to a transformation of $\pi$ and $\pi^{\ast}$
orbitals to $\sigma$ and $\sigma^{\ast}$ (see Fig. \ref{bands}).
The cohesive energy was found to be relatively small (of order 0.4
eV per hydrogen atom) which allows to make the process of
hydrogenation reversible \cite{H}. Roman and coworkers
\cite{jap1,jap2} considered different configurations of hydrogen
on graphene for the case of one-side hydrogenation and have found
that configurations when hydrogen atoms are bonded with different
sublattices (similar to Fig. \ref{sublatt}c) are the most stable.
In our work \cite{H} some general principles of hydrogenation of
graphene have been formulated based on calculations for various
configurations in a broad range of coverage.

\begin{figure}
\includegraphics[width=3.2 in]{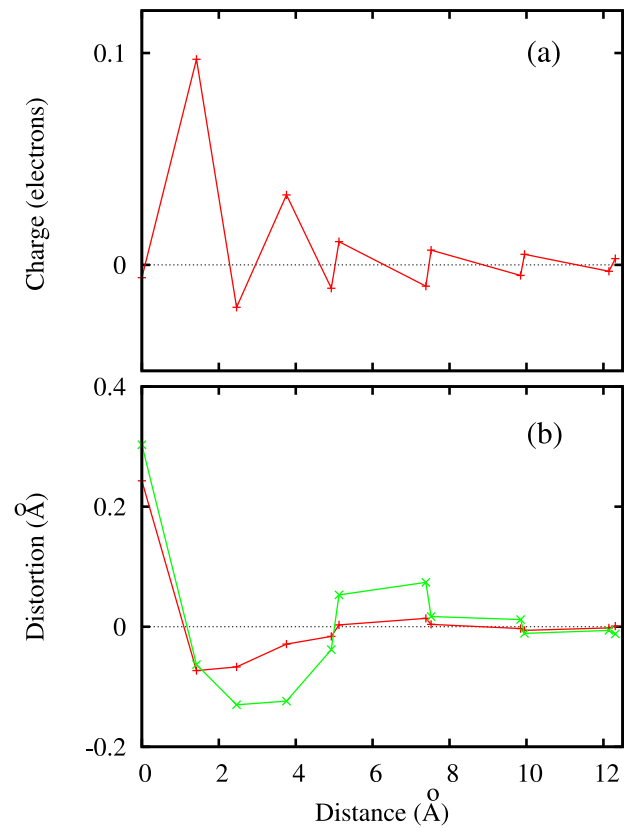}
\caption{Charge distribution for unpaired electron as a function
of a distance from the central carbon atom where hydrogen is
chemisorbed (a) and deviations of carbon atoms from flat
configuration for the cases of single hydrogen atoms (red solid
line) and two atoms chemisorbed on neighboring sites from
different side of the graphene sheet (green line) (b).
\label{dist} (color online)}
\end{figure}

Carbon atoms in graphene are in the sp$^2$ hybridization state
when each carbon atom has three $\sigma$ and one $\pi$ orbitals
(Fig. \ref{bands}). In contrast with numerous polyaromatic
hydrocarbons with localized single- and double bonds
$\pi$-orbitals in graphene are delocalized and all conjugated
chemical bonds are equivalent. Chemisorption of hydrogen atom
means a break of one of the $\pi$ bonds and transformation from
sp$^2$ to sp$^3$ hybridization. At the same time, unpaired
electron, one of those forming the $\pi$-bond, remains at the
neighboring carbon atoms. This electron is smeared in one of the
sublattices (Figs. \ref{dist}b and \ref{dist}a) and forms a
magnetic moment. The distribution of spin density in a radius of 1
nm correlates with distortions of the crystal lattice (Fig.
\ref{dist}b). However, this situation is not robust in a sense
that the chemisorption of the next hydrogen atom bonds this
unpaired electron and kills the magnetic moment \cite{H}.

Thus, the first principle of the chemical functionalization is the
absence of unpaired electrons, or, in other words, of dangling
bands. As a consequence of this principle, chemisorption of
functional groups on different sublattices are much more
energetically favorable than on the same sublattice.

\begin{figure}
\includegraphics[width=3.2 in]{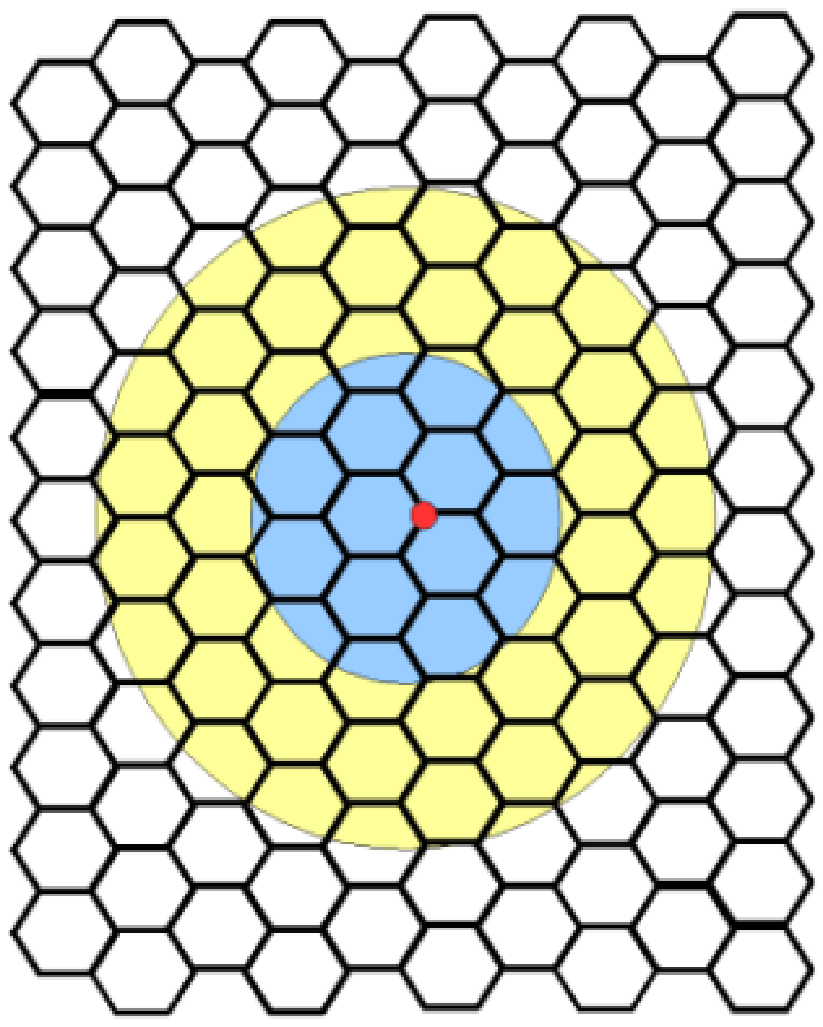}
\caption{Regions of strong (blue circle) and weak (yellow circle)
distortions of graphene lattice at the chemisorption of single
hydrogen atom (small red circle). \label{radii} (color online)}
\end{figure}

The next principle is a minimization of geometric frustrations.
One can see in Fig. \ref{dist}b that the chemisorption of single
hydrogen atom results in essential atomic displacements inside a
circle of radius 5 \AA~ (two periods of graphene crystal lattice)
and in smaller but still noticeable distortions with a
characteristic radius 10 \AA~ (Fig. \ref{radii}). The
chemisorption of single hydrogen atom leads to a strong shift up
of the carbon atom bonded with the hydrogen and shifts down for
two neighboring atoms. It is worth to notice that if one fixes
positions of all carbon atoms except nearest and next-nearest
neighbors of the central atom \cite{france} it changes essentially
the picture of atomic displacements. Such a procedure may be
reasonable to simulate chemical functionalization for constrained
systems such as graphene on graphite \cite{Andrei}, Ru
\cite{Ru1,Ru2,Pt}, Ir \cite{Ir} and Pt \cite{Pt} where part of
graphene lattice is strongly coupled with the substrate.

Chemisorption of the next hydrogen atom is more favorable when it
results in minimal additional distortions. This means that the
most energetically stable configuration arises if two hydrogen
atoms are chemisorbed by neighboring carbon atoms at different
sides of graphene sheet. One can see in Fig. \ref{dist}b that in
this case no essential additional atomic shifts are necessary. If
only one side is available for the chemisorption the most
favorable configuration of two hydrogen atoms is the bonding with
carbon atoms in the opposite corners of the hexagon in honeycomb
lattice (cites 1,4, or para-position, according to the chemical
terminology) \cite{bilay}. For the case of two more distant
hydrogen atoms they should migrate to this optimal position
overcoming some potential barriers \cite{move} which was observed
experimentally for both graphite \cite{HOPG} and graphene
\cite{Meyer}.

These two principles of the chemical functionalization predict
that in the most stable configuration for hydrogen pair on
graphene the hydrogen atoms will sit on neighboring carbon atoms
at the opposite sides with respect to the graphene sheet. Due to
geometric distortions a region of radius 5 \AA~ around the pair is
more chemically active than pure graphene, similar to the case of
graphene with defects \cite{Leht,swss,B,N,SW} and, thus, next
hydrogen atoms will be chemisorbed near the pair. In the case of
graphene with defects this process involves some potential
barriers whereas for ideal infinite graphene it will proceed
without any obstacles, up to complete hydrogenation and formation
of graphane.

In further works \cite{Yakobson,france} the results \cite{H} on
stability of nonmagnetic pairs of hydrogen on graphene and on the
formation energies have been confirmed, also the case of
polyaromatic hydrocarbons C$_{54}$H$_{18}$ \cite{Yakobson} and
C$_{42}$H$_{16}$ \cite{france} and C$_{96}$H$_{24}$ \cite{it}
being considered. Recent paper \cite{Blase} dealing with
adsorption and desorption of biphenyl on graphene demonstrates
importance of step by step simulation of the chemical reaction.

It is instructive to compare peculiarities of chemical
functionalization of graphene and carbon nanotubes. Similar to
graphene, in nanotubes the most favorable place for chemisorption
of the second hydrogen atom belongs to the opposite sublattice
with respect to the first one \cite{KrashNJP}. There is no direct
experimental evidences of this structure for the case of hydrogen,
however, for another functional groups it is known that normally
they are attached to pairs of neighboring carbon atoms
\cite{NTchem,cntf3,cntf4}. Minimization of geometric frustrations
plays an important role also in the case of nanotubes. In contrast
with flat graphene, a particular shape of the nanotubes makes the
chemisorption from external side more favorable, with the adsorbed
pairs situated along the nanotube. For large enough groups they
are usually attached to the opposite sides of the nanotube cross
section \cite{NTchem}, again, to minimize the carbon net
distortions. In contrast with graphene where complete {\it
one-side} hydrogenation is unfavorable for the case of nanotubes
it turns out to be possible, from external side, as is shown both
theoretically \cite{Dragan} and experimentally
\cite{Dillon1,Dillon2}.

\section{Functionalization of graphene  by other chemical species}

As discussed above, graphane is not very stable which is actually
quite good from the point of view of hydrogen storage since it
allows to hydrogenate and dehydrogenate graphene at realistic
temperatures \cite{Hsubmit}. At the same time, for potential use of
graphane in electronics this can be considered as a shortcoming.
Therefore it is interesting to consider other functional groups to
search a compound with the electronic structure similar to
graphane but larger cohesive energy.

\begin{figure}
\includegraphics[width=3.2 in]{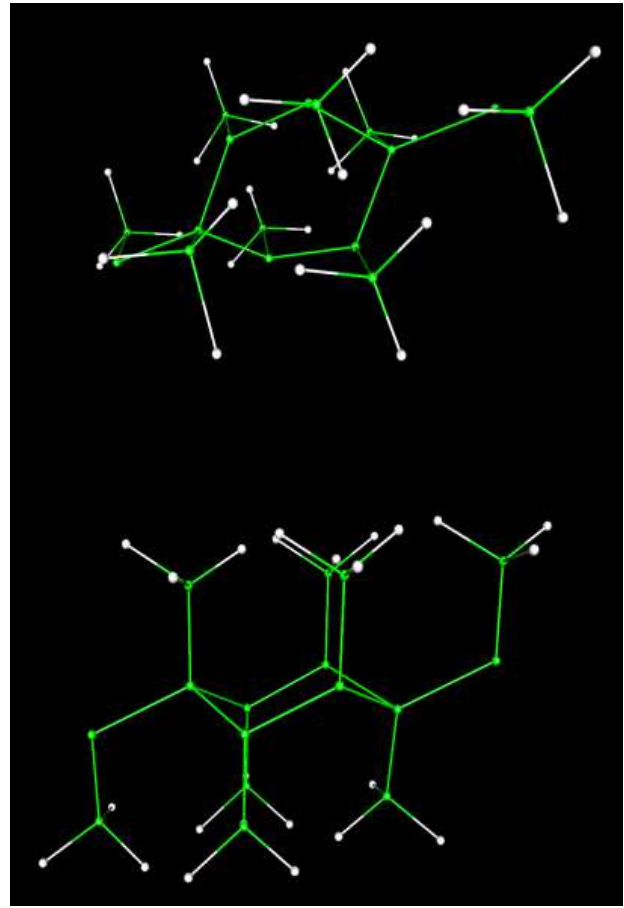}
\caption{Top and side view of graphene functionalized 
by CH$_3$ groups (optimized atomic positions).
Carbon and hydrogen atoms are shown in green and white, 
respectively.  \label{ch3} (color online)}
\end{figure}

The computational results are presented in Table \ref{tab1} and in
Fig. \ref{ch3}. Based on them one can divide all the functional
groups in two classes, (a) with the energy gap of order 3 eV and
large enough cohesive energy and (b) with much weaker bonding and
much smaller energy gap, or without a gap at all. This difference
can be related with the length of chemical bonds between carbon
atoms (carbon-carbon distance). For the substances from the first
class it is almost equal to that in diamond, 1.54 \AA~ whereas for
the substances from the second class it is larger. This means that
for the latter case, due to mutual repulsion of the functional
groups, graphene is destabilized which diminishes the cohesive
energy. This situation is observed experimentally in graphite
oxide which cannot be completely functionalized due to interaction
between hydroxy groups \cite{GO}, in agreement with the XPS data
\cite{GOxps}.

\begin{table}
\caption{\\ Dependence of the carbon-carbon bond length ($d$), in
\AA~, and the electron energy gap ($\Delta E$), in eV, on chemical
species for the case of 100\% two-side coverage of graphene.}
\label{tab1}
 \begin{center}
   \centering
\begin{tabular}{ccc}
\hline
Chemical species & $d$ & $\Delta E$ \\
\hline
 NO       & 1.622 & none \\
 NH$_2$   & 1.711 & 2.76 \\
 CN       & 1.661 & 2.11 \\
 CCH      & 1.722 & 0.83 \\
 CH$_3$   & 1.789 & 0.84 \\
 OH       & 1.620 & 1.25 \\
\hline
 H        & 1.526 & 3.82 \\
 F        & 1.559 & 4.17 \\
 H and F  & 1.538 & 5.29 \\
 H and Cl & 1.544 & 4.62 \\
 \hline
\end{tabular}
 \end{center}
\end{table}

As examples of chemical species which can provide, similar to
hydrogen, a complete coverage of graphene, we will consider F$_2$,
HF and HCl. Fluorine seems to be very promising for the
functionalization of graphene since it should produce a very
homogeneous structure, with a complete coverage by atoms of the
same kind. Therefore, this might be a way to create a
two-dimensional crystal with rather large energy gap but high
electron mobility, due to small degree of disorder. However,
fluorine is very toxic and very aggressive which may be a problem
for industrial use. Also, fluorine is used for ripping of
nanotubes \cite{ftor1,ftor2} which means that, potentially, edges
and defects of graphene lattice interacting with the fluorine can
be centers of its destruction. Actually, graphene samples are
always rippled \cite{rpl1,rpl2,rpl3,rpl30,rpl4,rpl5} or can be
deformed as observed for graphene ribbons \cite{cut3}.

\begin{figure}
\includegraphics[width=3.2 in]{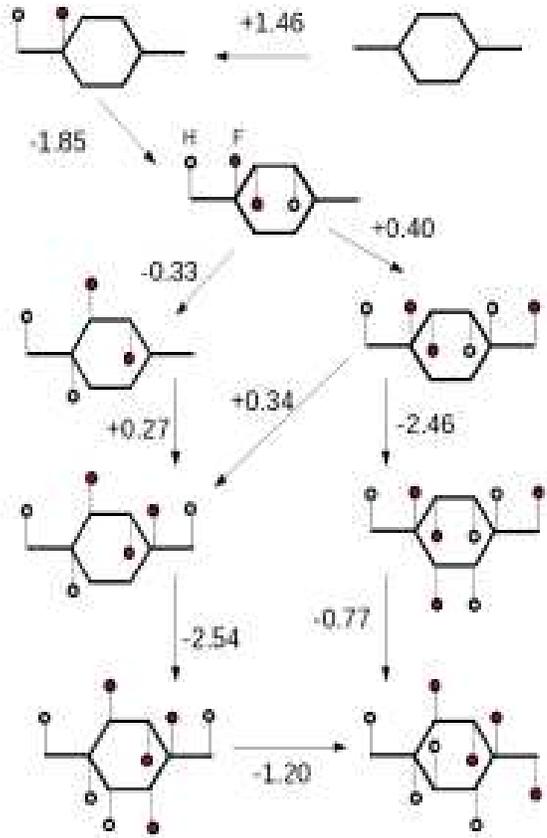}
\caption{A sketch illustrating processes and energetics of
different reactions of graphene with hydrofluoric acid. White and
red circles are hydrogen and fluorine atoms, respectively, all
energies are in eV. \label{HFbulk} (color online)}
\end{figure}

Let us consider now, step by step, interaction of graphene with
inorganic acids HF and HCl. The process of functionalization of
graphene by HF starts with the chemisorption of fluorine and
hydrogen atoms at neighboring carbon sites at the same side of
graphene sheet (Fig. \ref{HFbulk}).

As a first step of chemisorption of hydrofluoric acid molecule we
consider bonding of hydrogen and fluorine atoms with the
neighboring carbon atoms at the same side of graphene sheet (Fig.
\ref{HFbulk}a). In contrast with molecular fluorine where the
chemisorption energy is negative (see above) the energy of this
step turns out to be positive, +1.46 eV, mainly, due to strong
distortions of initially flat graphene at the one-side
functionalization. As the next step, we consider chemisorption of
one more H-F pair from another side of graphene (fig.
\ref{HFbulk}b). Similar to chemisorption of the second hydrogen
atom (see above) this step is energetically favorable, with the
formation energy -1.85 eV. Reconstruction to the structure shown
in Fig. \ref{HFbulk}c diminishes the total energy by 0.33 eV which
makes this a possible third step of the process under
consideration. After that, the chemisorption energy of third HF
molecule turns out to be much smaller than for the first one but
still positive, +0.27 eV (Fig. \ref{HFbulk}d). Starting from the
fourth molecule, the chemisorption energies are negative so
further reaction is exothermal. As a result, complete coverage of
graphene by hydrofluoric acid is energetically favorable but
requires high enough energy for the first step. This means that,
the most probably, bulk graphene without defects will be stable
enough with respect to reaction with HF, at least, at room
temperature and normal pressure. It seems to be practically
important since hydrofluoric acid is used to solvate SiO$_2$
substrate when prepare freely hanged graphene membrane
\cite{Andrei}. One can expect in this case a formation of a
chemically modified layer near edges with essentially different
electronic structure (see below).

Our calculations show that the first step of functionalization of
graphene by hydrochloric acid is essentially different, due to
larger interatomic distance (1.27 \AA ~for HCl pair, instead of 0.97 \AA ~
for HF) which close to carbon-carbon distance in graphene (1.42
\AA). As a result, there is no break of bonds between hydrogen and
chlorine but, instead, HCl molecule hangs over graphene layer
bonding with it by only weak van-der-Waals bond and there is no
chemical reaction. Thus, the hydrocloric acid can be safely used
to clean graphene.

\begin{figure}
\includegraphics[width=3.2 in]{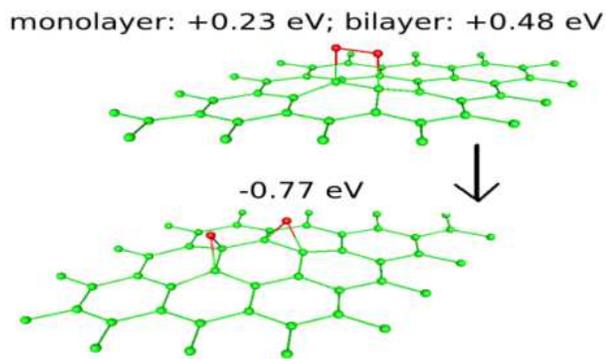}
\caption{Optimized geometric structure for step-by-step adsorption
of oxygen molecule on graphene. \label{oxo} (color online)}
\end{figure}

Let us consider now, using oxygen molecule as an example, a
chemisorption of diatomic molecule with {\it double} chemical
bond. In this case, at the first step the double bond will
transform into the single one, with formation of single bonds with
neighboring carbon atoms (Fig. \ref{oxo}). This reaction requires
to overcome some energy barrier which is, however, rather low
since flexibility of graphene allows to minimize the energy costs
due to lattice distortions. Bilayer graphene is much less flexible
\cite{bilay}, due to interlayer coupling, which makes the
chemisorption energy higher, in agreement with the experimental
data \cite{Flynn}.

If we would calculate chemisorption energy with respect to not
singlet but triplet state of O$_2$ molecules it would increase all
chemisorption energies by 1.12 eV (see Section 2). The resulting
energies are so high that it seems in contradiction with the
experimental fact that graphene is relatively easily oxidized
already at temperature up to 200 \textcelsius \cite{Flynn}. Also, ``unzipping''
process at the oxidation of graphite observed experimentally
\cite{unzip,Ogr} will involve energetically unstable states.
Graphene oxide turns out to be also unstable with respect to
triplet oxygen \cite{GO,reduct2}. Probably these discrepancies mean
that formation of singlet oxygen plays some role in these
processes. This issue requires further investigation. A similar
problem was discussed earlier for nanotubes \cite{O1}.

Oxidation energy for graphene turns out to be lower than for the
case of nanotubes \cite{O1,O2a}. In the latter case, $\pi$-orbitals are
rotated one with respect to another \cite{cntf1} which makes
formation of oxygen bridges between carbon atoms more difficult.

Previous computational results for oxidation of graphene
\cite{Oribb} are qualitatively similar to ours. However, the energy
of cyclo-addition there is higher than ours. There are two
possible reasons for this difference, the use of LDA in Ref.
\cite{Oribb} instead of GGA in our calculation and the use of
relatively small graphene ``molecule'' (with 64 atoms) instead of
infinite system with periodic conditions here.

\begin{figure}
\includegraphics[width=3.2 in]{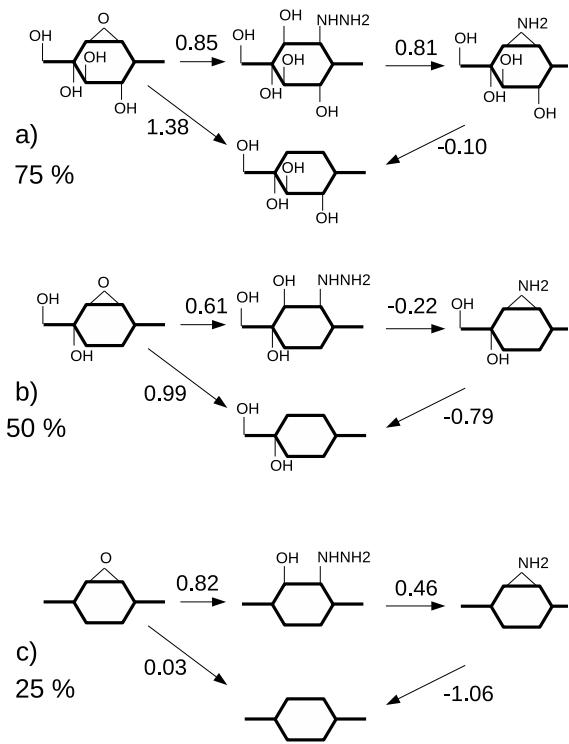}
\caption{Sketch of chemical reduction of graphene oxide
\cite{reduct2} for various degrees of its functionalization. All
energies are in eV. \label{GOreduct} (color online)}
\end{figure}

Let us consider now another important issue, the simulation of
reduction of graphite oxide (GO). We have proposed earlier
\cite{GO} models of this compound with different degrees of its
reduction. In ref. \cite{reduct2} a process of the reduction has
been studied and the scheme of corresponding reactions has been
proposed. Using our model and this scheme, we have simulated three
possible ways of the chemical reduction of GO depending on degree
of preliminary reduction and calculated the corresponding energies
(Fig. \ref{GOreduct}). To compare, we have calculated also energy
costs of direct reduction (e.g., by heating). One can see from the
Figure that for high degrees of coverage of GO (weakly reduced)
energy costs of chemical reactions are smaller than those of the
direct reduction whereas for the case of strongly reduced GO, vice
versa, direct total reduction turns out to be more energetically
favorable. Similar, other reactions of real \cite{polymer,Si} and
potential functionalization of GO can be simulated which may be
important to search ways of its complete reduction to pure
graphene.

\section{Functionalization of graphene nanoribbons}

\begin{figure}
\includegraphics[width=3.2 in]{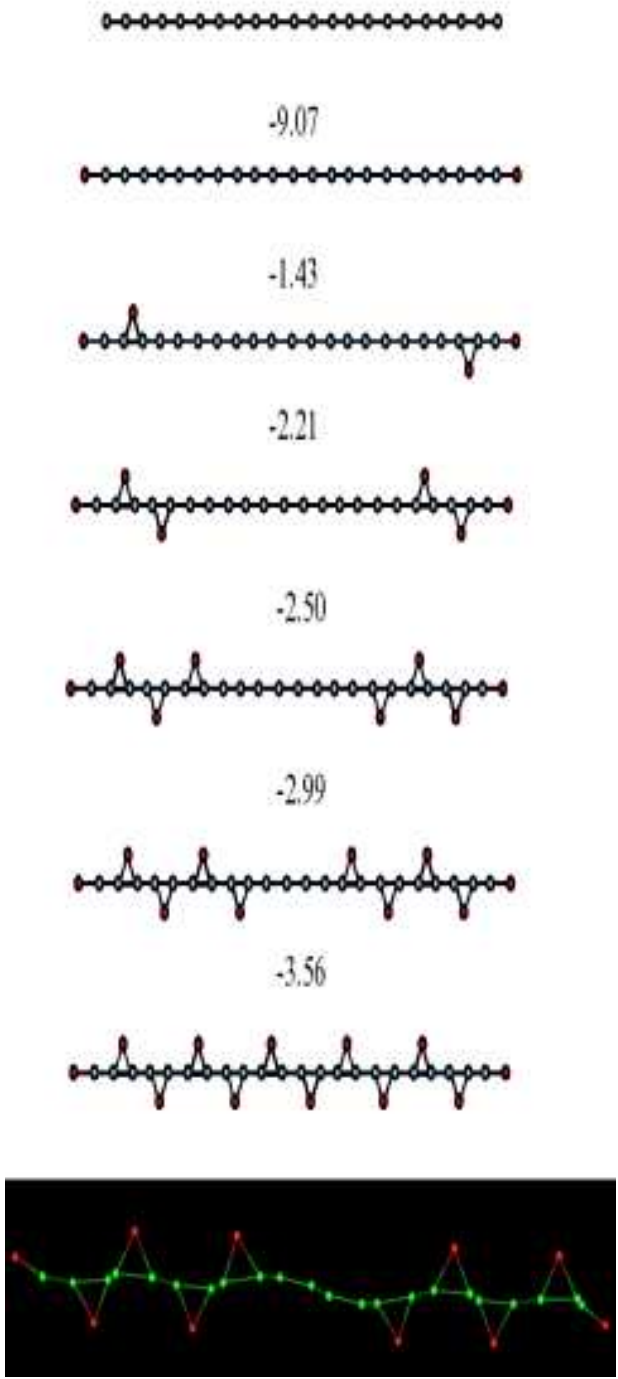}
\caption{A scheme of step-by-step oxydation of GNR of width 2.2
nm. All energies are in eV (top panel).
Optimized geometric structure for partialy oxidized
GNR (lower panel). \label{Orib22} (color online)}
\end{figure}

\begin{figure}
\includegraphics[width=3.2 in]{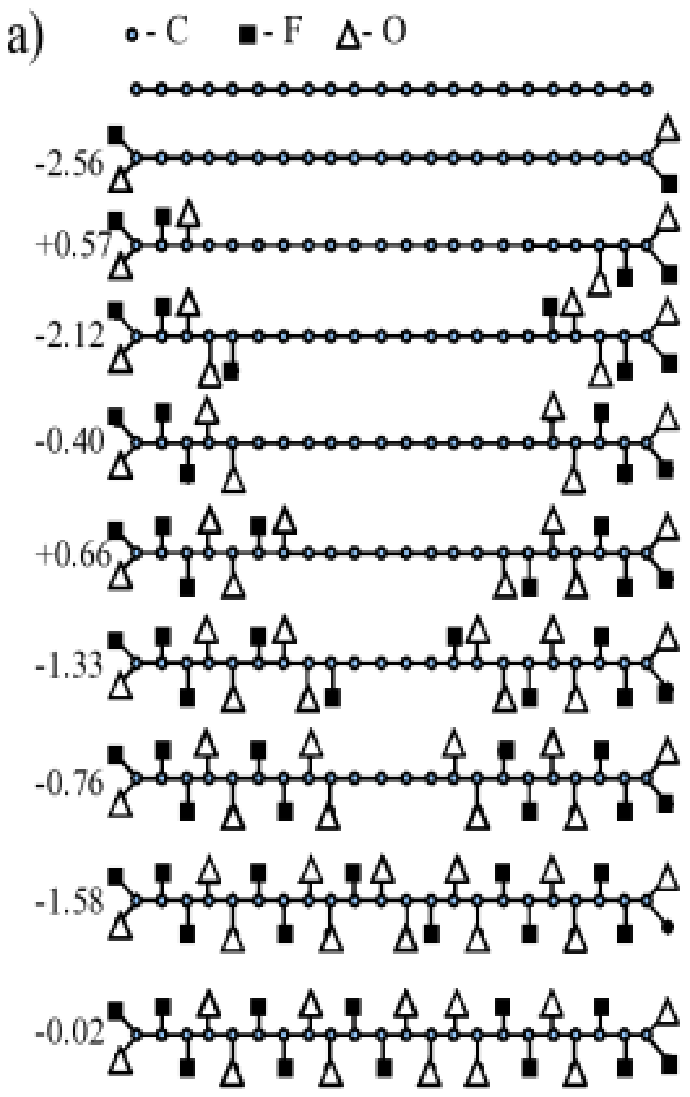}
\rotatebox{-90}{
\includegraphics[height=3.2 in]{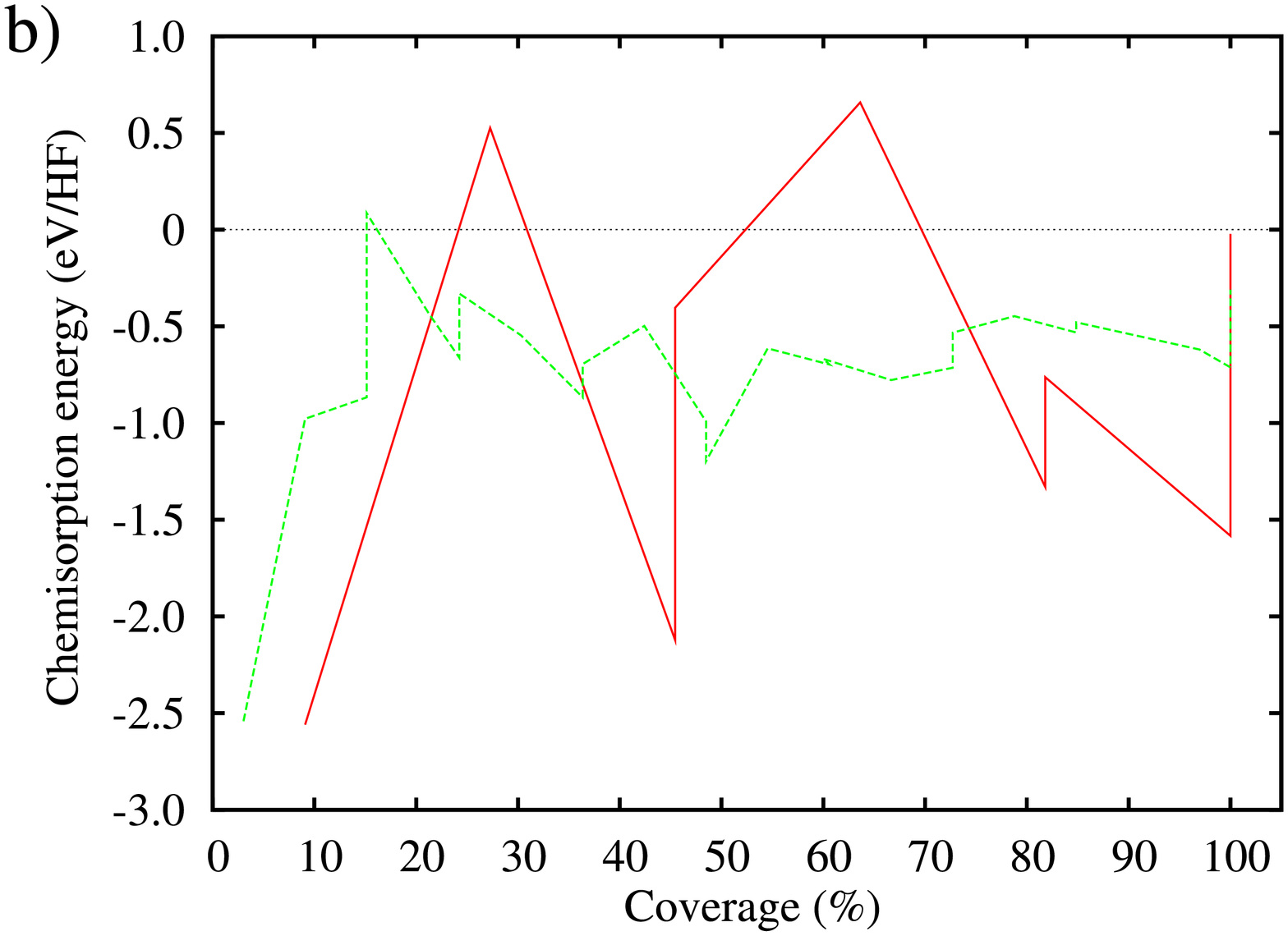}}
\caption{A scheme of step-by-step reaction of GNR of width 2.2 nm
with HF; the energies are in eV (a). The same energies as
functions of coverage for the case of GNR with width of 2.2 nm
(red solid line) and 6.6 nm (green dashed line). \label{HF22}
(color online)}
\end{figure}

Polyaromatic hydrocarbons (PAH) \cite{chemgraph} are quite close
to graphene nanoribbons (GNR), and their chemistry, in particular,
hydrogenation of their edges, is rather well studied
\cite{2bonds}. Another nanoobjects which can be considered as
predecessors of GNR are nanographite and nanographite intercalates
which were studied, in particular, in connection with their
magnetism \cite{Enoki1,Enoki2}. Model calculations of their
magnetic properties are presented in Refs. \cite{Enoki1,Enoki2}
whereas first-principle calculations, using a fragment of graphene
sheet as a model of nanographite, have been performed in Ref.
\cite{nanograph}. In that paper, functionalization of the graphene
edges by single hydrogen atom has been considered, as well as
suppression of magnetism by fluorination, in relation with the
experimental data \cite{EnokiF}.

Starting from the first work by S. Louie and coworkers
\cite{Louie} predicting, on the base of the DFT calculations,
half-metallic ferromagnetic state of zigzag graphene edges, this
problem attracts a serious attention \cite{YK,kit2}. Later, the
first results have been reconsidered by the same group using more
accurate many-body GW method \cite{Yang}. Chemical
fucntionalization of GNR is currently one of the most popular
subjects in graphene science.

O. Hod et al. have considered effects of the shape of
semiconducting GNR on their stability and electronic structure
\cite{Hod1}, electronic structure of graphenic nanodots
\cite{Hod2} (similar calculations of magnetism and electronic
structure have been presented also in Ref. \cite{magnonano}),
effects of geometry of graphene nanosheets on their electronic
structure \cite{Hod3} and enhancement of half-metallicity due to
partial oxidation of the zigzag edges \cite{Hod4}. Interesting
effects which can arise in graphene nanoflakes are discussed in
Ref. \cite{Kaxiras}. Usually, it is supposed in these calculations
that edges of nanoflakes are passivated by hydrogen atoms
\cite{Hod2,Hod3,Kaxiras,fl1,fl2,fl3} but sometimes pure edges are
considered \cite{fl4,fl4}. Real structure of some graphene
nanoflakes was discussed in Ref. \cite{radovic}. In general, the
state of real graphene edges is not well understood yet and this,
very important, problem needs further investigations.

In Ref. \cite{Huang} effect of vacancies at the edges on the
electronic structure of GNR were studied. Influence of
chemisorption of different chemical species on the electronic
structure of GNR was investigated in Refs. \cite{Ferarri2,kit1}.
Ref. \cite{O2} considered the chemisorption of molecular oxygen on
GNR and its effect on their magnetism. Various configurations of
hydrogenated \cite{Mauri} and self-passivated \cite{fin} GNR were
studied; in both cases nonmagnetic state of the GNR turned out to
be the most stable. Refs. \cite{SW,White} demonstrated suppression
of magnetism at the zigzag edges by oxidation. Since the latter is
a natural process at the production of GNR \cite{cut3,cut4,dress}
some special measures to protect the magnetism chemically are
probably necessary.

Thus, works on the functionalization of GNR can be divided into
three groups. First, there are investigations of electronic
structure and physical properties (especially, magnetism) of the
edges themselves. This group includes most of theoretical works.
Second, there are works studying the functionalization of the
nanoribbons between the edges. It seems that the effect of the
edges on the electronic structure of GNR is long-range enough and,
as a consequence, chemistry of GNR can be essentially different
from that of bulk graphene. Research in this direction is just
starting. Third, there are attempts to find optimal geometric and
chemical structure of GNR taking into account realistic
technological processes of their formation. Preliminary, there is
an impression that some structures considered theoretically before
can be very different from the real ones, so we are, indeed, only
in the beginning of the way. On the other hand, a lot of
experimental and theoretical information is available for related
compounds such as arynes \cite{aryne}, polyhexacarbens
\cite{hcarb} and polycyclic aromatic hydrocarbons \cite{PAH2}
which may be also relevant for GNR.

It was shown in our previous work \cite{SW}, using hydrogen as an
example, that the edges of GNR are centers of chemical activity
and their functionalization can be just a first step to the
functionalziation of the whole nanoribbon. Here we will consider
step by step the oxydation process of GNR. The computational
results are presented in Fig. \ref{Orib22}. Thus, oxidation of GNR
is a chain of exothermal processes. This result may be relevant
also for better understanding of burning and combustion of carbon
systems. These issues, as well as more detailed results concerning
oxidation and unzipping \cite{unzip} of graphene and related
compounds will be discussed elsewhere.

Let us consider now a reaction of GNR with the hydrofluoric acid.
A scheme of this reaction for the nanoribbon of width 2.2 nm is
shown in Fig. \ref{HF22}a and dependence of the energy as a
function of coverage is presented in Fig. \ref{HF22}b. In contrast
with the case of ideal infinite graphene for narrow enough GNR
{\it two} steps with positive chemisorption energy. This energy
cost is so small that the process of hydrofluorination of this
nanoribbon can take place even at room temperature. The situation
is essentially different for the nanoribbon of width 6.6 nm. The
steps with positive chemisorption energy just disappear, due to
larger distance between the regions of the reaction. Thus, the
hydrofluorination of a broad enough GNR is easy and complete.

One can see, that, similar to the case of hydrogenation \cite{SW}
(see also the previous section) GNR are very chemically active
which probably means that, at least, some of their applications
will require an inert atmosphere.

\section{Conclusions and perspectives}

In Section 3 we discussed, using hydrogenation as example, three
main principles of chemical functionalization of graphene: (i)
broken bonds are very unfavorable energetically and, therefore,
magnetic states on graphene are usually very unstable; (ii)
graphene is very flexible, and atomic distortions influence
strongly on the chemisorption process; (iii) the most stable
configurations correspond to 100\% coverage for two-side
functionalization \cite{H} and 25\% coverage for one-side
functionalization \cite{bilay}. Based on these principles one can
model the functionalization of not only perfect graphene
\cite{H,bilay,GO} but also graphene with intrinsic and extrinsic
defects which are centers of chemical activity \cite{SW}.

In Section 4 we have studied chemical functionalization of
graphene by fluorine and hydrofluoric acids. They can provide
complete coverage and, thus, semiconducting state with large
enough electron energy gap and weak disorder. Alternatively,
non-covalent functionalization can be used to create the gap.

The current situation seems to be a bit controversial. In some
papers \cite{CrO3,Peres} an energy gap opening due to
physisorption of various molecules has been found, in contrast
with other results \cite{Tim2,Tim3,ahuja1,antwerp,h2so4}. It is
difficult to compare these works directly since they are, in
general, done for different chemical species and their
concentrations. However, keeping in mind that in some cases
\cite{Tim3,antwerp,Peres} the same case of water has been studied
one can conclude that the electronic structure is very sensitive
to specific concentration and specific geometric configuration of
water on graphene surface. At the same time, one should keep in
mind that a very specific structure of water \cite{Ad} and other
hydrogen-bonded substances may be not very accurately reproduced
in the DFT calculations. Probably, only combination of the DFT,
quantum chemical calculations and molecular dynamics can clarify
the situation.

Anyway, since the cohesive energy at physisorption does not exceed
20 kJ/mol the physisorbed graphene cannot be very robust which may
result in some restrictions of its use in electronics.

We have demonstrated using HF as an example that it is important
to simulate the chemisorption process step by step since it allows
to estimate accurately chemisorption energies relevant for each
step of the process. It turns out that for perfect infinite
graphene the energy cost of first step for the case of HF is
rather high whereas for HCl this reaction is practically
impossible. This means that these acids can be safely used for
cleaning of graphene samples. At the same time, the potential
barrier for the oxidation is rather low which makes even moderate
annealing of graphene in oxygen-contained atmosphere not very safe
procedure.

We have shown also, in the last section, that the graphene
nanoribbons are chemically active enough to be oxidized and to
interact with hydrofluoric acid with relatively low activation
energy. It means that to prepare the graphene nanoribbons with a
given chemical composition alternative ways should be used, e.g.,
its synthesis from polyaromatic hydrocarbons \cite{chemgraph,
radovic} or even use microorganisms as it is used already for
biogenic grow \cite{biogrow1,biogrow2} and cleaning
\cite{daphnia}.

\section*{Acknowledgements}

We are grateful to A. K. Geim, K. S. Novoselov, E. Y. Andrei, A.
V. Krasheninnikov, A. I. Lichtenstein, T. Wehling, P. A. Khomyakov
and O. V. Yazyev for helpful discussions. The work is financially
supported by Stichting voor Fundamenteel Onderzoek der Materie
(FOM), the Netherlands.

\end{document}